\documentstyle[twoside,fleqn,espcrc2]{article}
\newcommand{\be}{\begin{equation}}
\newcommand{\ee}{\end{equation}}

\newcommand{\AmS}{{\protect\the\textfont2
  A\kern-.1667em\lower.5ex\hbox{M}\kern-.125emS}}

\title{Extreme Energy Cosmic Rays (EECR) Observation 
Capabilities of an ``Airwatch from Space'' Mission}
\author{presented by C. N. De Marzo\address{Dipartimento 
di Fisica, Universit\`a e Sezione INFN, Bari, Italy,
e-mail: demarzo@ba.infn.it} 
for the Airwatch Collaboration$^{(*)}$}
\begin{document}
\begin{abstract}
The longitudinal development and other characteristics of the 
EECR induced atmospheric showers
can be studied from space by detecting the fluorescence light induced in the 
atmospheric nitrogen. 
According to the Airwatch concept a single fast 
detector can be used for measuring both intensity and time development of the 
streak of fluorescence light produced by the atmospheric shower induced by an 
EECR. 
In the present communication the 
detection capabilities for the EECR observation from space are discussed.
\end{abstract}
\maketitle
\section{The Airwatch from space mission}
EECR with energies $> 10^{20}$ eV 
have been detected through the 
giant showers they produce in the atmosphere
\cite{EECR}. Their existence raises 
questions on their origin and propagation mechanism above
the `Greisen-Zatsepin-Kuzmin' cutoff ~\cite{Greisen66} in the energy spectrum
at $\sim 6 \times 10^{19}$ eV.
Because of their very high energy, if EECR point to definite sources,
an explanation in terms of a ``bottom-up'' acceleration mechanism
would be favoured.
On the other hand, if they indicate broader or structured regions, 
a ``top-down'' model considering them as decay products 
of topological defects should be taken into account~\cite{Sigl95}. 
The EECR shower development is accompanied by UV emission, in 
particular by the fluorescence light induced in the atmospheric nitrogen which 
has characteristic spectral lines in the near UV. The possibility to detect 
air showers through the fluorescence light they produce has 
been proven by detectors like Fly's Eye~\cite{Baltrusaitis85}, 
whose operation has given information on flux, incoming direction and 
composition of the primary particles. 

An atmospheric mass 2-3 orders of magnitude larger can be monitored 
by detecting the 
fluorescence light induced by EECR in the atmosphere from a space 
platform~\cite{Linsley79}. 
The proposed ``Airwatch'' Mission appears to have the capability of
measuring the EECR spectrum above $10^{20}$ eV, by
determining the arrival direction, the energy, the 
$X_{max}$ and the primary species for each shower.
This capability appears to provide a 
complemental approach to large ground arrays and promises the possibility 
of extending EECR 
measurements beyond $10^{21}$ eV~\cite{Takahashi95,Barbier96}.

The amount of atmospheric area and mass that can be observed 
from a space platform with reasonable design parameters --- 
orbit height and field of view (FOV) --- is given in Tab.~\ref{tab1}.
\begin{table*}[htb]
\setlength{\tabcolsep}{.5pc}
\newlength{\digitwidth} \settowidth{\digitwidth}{\rm 0}
\catcode`?=\active \def?{\kern\digitwidth}
% -----------------------------------------------------
\caption{Atmospheric area and mass as a function of orbit height and FOV.}
\label{tab1}
\begin{tabular*}{\textwidth}{@{}c@{\extracolsep{\fill}}lll}
\hline
                 \multicolumn{1}{c}{Orbit height (km)}
                 & \multicolumn{3}{c}{FOV (degrees)}\\
\cline{2-4}
& \multicolumn{1}{c}{$60^{\mbox{o}}$} &\multicolumn{1}{c}{ $90^{\mbox{o}}$} 
& \multicolumn{1}{c}{$120^{\mbox{o}}$} \\ \hline
400 & $0.17 \times 10^{6}$ km$^{2}$ & $0.50 \times 10^{6}$ km$^{2}$ & 
$1.5 \times 10^{6}$ km$^{2}$\\ 
    & $1.7 \times 10^{12}$ ton & $5 \times 10^{12}$ ton & 
$15 \times 10^{12}$ ton\\ \hline
500 & $0.26 \times 10^{6}$ km$^{2}$ & $0.80 \times 10^{6}$ km$^{2}$ &
$2.4 \times 10^{6}$ km$^{2}$\\
    & $2.6 \times 10^{12}$ ton & $8 \times 10^{12}$ ton & 
$24 \times 10^{12}$ ton\\ \hline
600 & $0.38 \times 10^{6}$ km$^{2}$ & $1.1 \times 10^{6}$ km$^{2}$ &
$3.4 \times 10^{6}$ km$^{2}$\\
    & $38 \times 10^{12}$ ton & $11 \times 10^{12}$ ton & 
$34 \times 10^{12}$ ton\\ \hline
\end{tabular*}
\end{table*}
It should be considered that the atmospheric mass observed by
Flye's Eye is of the order of $10^{9}$ tons.

The fluorescence light produced by an atmospheric shower of extreme 
energy is mainly emitted in the near UV molecular nitrogen lines 337, 357 and 
391 nm. As seen from space, in these spectral bands the shower appears as a 
spot of light travelling with speed near to that of light and with changing 
intensity. The fluorescence light is accompanied by Cherenkov light, 
by Rayleigh and Mie scattered Cherenkov light. All these components may
be reflected on clouds or ocean surface (`end point flash'). 
These optical phenomena bring further information on the shower 
development. 
The spot velocity, projected on the horizontal plane, equals the velocity 
of light times the cosine of the angle of its direction with the horizontal 
plane. This fact provides one possible approach to measure the 
direction of the shower axis: i.e. by measuring the velocity position 
and time of the pixels fired in the focal plane detector. 

In an alternative approach 2 focal plane detectors on 2 separate 
platforms, orbiting 600 km apart, are used for stereoscopic view -
OWL concept~\cite{Barbier96}. 
%\section{Detector features and counting rate}

Assuming that the shower streak of light is long 10 to 100 km in the 
atmosphere, the Airwatch focal plane detector segmentation must be of the 
order of $1000 \times 1000$ pixels for monitoring one million km$^{2}$ 
of atmosphere with sufficient resolution and counts per year. 
Assuming a 5 m$^{2}$ light collector (1.3 m radius) 
and resonable values for its optical efficiency; considering that fluorescence 
light yield, in the 300--400 nm spectral interval, is $\sim 4$ $photon/
(particle \cdot$ m) and that a $10^{20}$ eV shower has $6 \times 10^{10}$ 
particles at its maximum development, 
the detector sensitivity has to work at a level of 10--100 $photons/pixel$ on 
integration times of about 3 $\mu$s. These requirements 
ask for a detector technology based on photon-electron 
conversion on a photocathode and subsequent electron multiplication.

The effective counting rate for an Airwatch Mission depends on the 
amount of watched atmosphere and on the shape of the energy spectrum. 
On the higher energy side 
counting rate is limited by the EECR low flux. 
On the lower energy side the detection efficiency depends on the trigger 
sensitivity.
The trigger electronics of an Airwatch from Space Mission can take 
advantage of the unique features of the showers: 
there is no possible background 
looking like a luminous spot moving at light speed through the atmosphere over 
tens of kilometers. The Airwatch detector and its trigger need to be 
smart in detecting showers on the lower energy side.  
At present stage of the project an energy threshold something 
below $10^{20}$ eV appears reachable, 
allowing the comparison with measurements of ground based detectors and arrays. 
For an atmospheric mass $\sim 10^{12}$ tons, an
order of 100 events per year could be collected above $10^{20}$ eV.

Operations with a $10 \div 20\%$ duty cycle are predicted for an Airwatch 
Space Mission, taking into account sunlight, moonlight, civilization 
ground lights, high clouds and average weather patterns. An equatorial orbit 
will maximize observation time over oceans and deserted areas. Radiation 
degradation of the optics is also minimized. Conversely, one drawback of an 
equatorial orbit is the higher cloud coverage. 

To pursue a complete study of all the sources of background --- meteors, 
aurorae, lightining --- foreseen for the experiment, 
a precursor measurement of the light phenomena in the atmosphere 
(EAGLE Mission) has been proposed~\cite{DeMarzo97}.
\section{On neutrino and GRB detection}
In a ``top-down'' acceleration scenario, a sensible flux of $\nu$'s 
is expected on the Earth up to EECR energies.  
Airwatch from Space can detect them
because of the huge and transparent target ($\sim 10^{12}$ tons) 
available for interaction and detection. Moreover, it has been shown 
that the cross section for $\nu$ interaction will increase with energy 
up to the order of $10^{2}$ nb at $10^{20}$ eV~\cite{Gandhi96}. 
Because atmospheric thickness rises of about a factor 40 as shower 
direction goes from vertical to horizontal angle, EECR showers initiated by 
neutrinos can be identified by their appearance at large zenith angles and 
large atmospheric depths where no hadron or photon initiated showers can be 
present.

While EECR showering in the atmosphere will appear as UV tracks 
spatially well defined and short in time, a GRB will arrive 
as a plane wave investing the entire atmosphere and exciting by Compton 
electrons a diffuse fluorescence emission having a peculiar space-time 
function~\cite{Catalano97}. This signal can be detected by Airwatch and 
possible correlations between EECR and GRB's can be studied. 
\section{Discussion and prospects}
The main design parameter for an Airwatch experiment is the height of 
the orbit.
At fixed FOV, on one hand the 
monitored atmospheric mass increases with the orbit altitude; on the other 
hand the signal strength decreases. The optimum solution must be decided 
through an effective simulation of the detector performance.

A possible approach could be based on 
the segmentation of the mission in a certain number of small satellites. 
This scenario would in general be cheaper and have a faster schedule than 
the traditional single or double observatory, presenting the possibility to 
be implemented with the methodology used in manufacturing large number of 
telecommunication satellites.

This solution 
can be approached through an exploratory mission on a single small satellite 
that will be very useful both from the engineering and the 
physical point of view. 
A spacecraft in the range of 400 kg and 400 W can have a collection 
capability of some tens of events per year, still providing very interesting 
results for the EECR study.
Finally, the possibility to locate an Airwatch from Space Mission on the 
Space Station would offer several technical advantages,
not least the opportunity to use a 9 m diameter collecting mirror whose 
technology is available~\cite{Garipov}. 
%by the russian colleagues of this collaboration~\cite{Garipov}.

(*) {\bf The Airwatch Collaboration:}\\
{ \footnotesize
M. Ambriola, R.Bellotti, F. Cafagna, F. Ciacio, M. Circella, 
C.N. De Marzo, N. Mirizzi, T. Montaruli - 
{\it Fisica, Universit\`a e Sezione INFN, Bari, Italy.}\\
G. Giovannelli, I. Kostandinov - {\it FISBAT-CNR, Bologna, Italy.}\\
G. Bonanno - {\it Oss. Astrofisico, Catania, Italy}, 
R. Fonte - {\it Sezione INFN, 
Catania, Italy and 
%Phys. Dep., 
Univ. of New Mexico, Albuquerque, USA.}\\
O. Adriani, G. Becattini, P. Spillantini - 
{\it Fisica, Universit\`a di Firenze, Italy.}\\
P. Mazzinghi, G. Toci -
{\it IEQ-CNR, Firenze, Italy.}\\
F. Fontanelli, V. Gracco, A Petrolini, G. Piana, M. Sannino -
{\it Fisica, Universit\`a e Sezione INFN, Genova, Italy.}\\
G. D'Al\`{\i}, L. Scarsi - 
{\it Energ. ed Appl. Fisica, Universit\`a 
di Palermo, Italy.}\\
G. Agnetta, O. Catalano, S. Giarrusso, M.C. Maccarrone, B. Sacco - 
{\it IFCAI-CNR, Palermo, Italy.}  \\
P. Lipari - {\it Sezione INFN, Roma, Italy.}\\
M. Stefani - {\it Fisica, Universit\`a di Roma 3, Italy.}\\
G. Giannini - {\it Fisica, Universit\`a di Trieste, Italy.}\\
V. Bratina, A. Gregorio, R. Stalio, P. Trampus, B. Visintini - 
{\it CARSO, Trieste, Italy.}\\
C. Cepek, A. Laine, E. Mangano, M. Sancrotti -
{\it Laboratorio TASC-INFM, Trieste, Italy.}\\
J.N. Capdevielle - {\it L.P.C. Coll\'ege de France, Paris.}\\
B. Khrenov, M. Panasyuk -
{\it Skobelisyn Institute of Nuclear Physics, Moscow State University, 
Russia.}\\
J. Linsley -
{\it 
%Phys. Dep., 
Univ. of New Mexico, Albuquerque, USA.}\\
Y. Takahashi -
{\it 
%Phys. Dep., 
Univ. of Alabama, Huntsville, USA.}\\
L.A. Broadfoot - {\it Lunar and Planetary Lab., Univ. of Arizona, Tucson, USA.}
}
\end{document}